\documentclass[iop]{emulateapj}
\slugcomment{Accepted for publication in ApJ: 2013 July 29}
\usepackage{graphicx}
\usepackage{aas_macros}
\usepackage{amsmath}
\usepackage{amssymb}
\newcommand{\dd}{\;\mathrm{d}}

\newcommand{\ergs}{$\mathrm{erg\;s^{-1}}$}
\newcommand{\msun}{\mathcal{M}_{\sun}}
\newcommand{\Msun}{$\msun$}
\newcommand{\mstel}{\mathcal{M}_*}
\newcommand{\Mstel}{$\mstel$}
\newcommand{\mbh}{\mathcal{M}_\mathrm{bh}}
\newcommand{\Mbh}{$\mbh$}

\newcommand{\lx}{L_\mathrm{X}}

\newcommand{\giv}{\;|\;}

\newcommand{\amend}[1]{{#1}} 
\newcommand{\moreamend}[1]{{#1}} 

\shorttitle{An observationally motivated model to connect the AGN and galaxy populations}
\shortauthors{Aird et al.}

\begin{document}

\title{PRIMUS: An observationally motivated model to connect the evolution of the AGN and galaxy populations out to $z\sim1$}
\author{
James Aird\altaffilmark{1,9}, 
Alison L. Coil\altaffilmark{2,10}, 
John Moustakas\altaffilmark{3},
Aleksandar M. Diamond-Stanic\altaffilmark{2,11},
Michael R. Blanton\altaffilmark{4}, 
Richard J. Cool\altaffilmark{5}, 
Daniel J. Eisenstein\altaffilmark{6}, 
Kenneth C. Wong\altaffilmark{7}, 
Guangtun Zhu\altaffilmark{8}
}
\email{E-mail: j.a.aird@durham.ac.uk}
\altaffiltext{1}{Department of Physics, Durham University, Durham DH1 3LE, UK}
\altaffiltext{2}{Center for Astrophysics and Space Sciences, Department of Physics, University of California, 9500 Gilman Dr., La Jolla, San Diego, CA 92093, USA}
\altaffiltext{3}{Department of Physics and Astronomy, Siena College, 515 Loudon Road, Loudonville, NY 12211, USA}
\altaffiltext{4}{Center for Cosmology and Particle Physics, Department of Physics, New York University, 4 Washington Place, New York, NY 10003, USA}
\altaffiltext{5}{MMT Observatory, 1540 E Second Street, University of Arizona, Tucson, AZ 85721, USA}
\altaffiltext{6}{Harvard College Observatory, 60 Garden St., Cambridge, MA 02138, USA}
\altaffiltext{7}{Steward Observatory, The University of Arizona, 933 N. Cherry Ave., Tucson, AZ 85721, USA}
\altaffiltext{8}{Department of Physics \& Astronomy, Johns Hopkins University, 3400 N. Charles Street, Baltimore, MD 21218, USA}
\altaffiltext{9}{COFUND Junior Research Fellow}
\altaffiltext{10}{Alfred P. Sloan Foundation Fellow}
\altaffiltext{11}{Center for Galaxy Evolution Fellow}

\begin{abstract}
We present an observationally motivated model to connect the AGN and galaxy populations at $0.2<z<1.0$ and predict the AGN X-ray luminosity function (XLF). 
We start with measurements of the stellar mass function of galaxies (from the Prism Multi-object Survey) and populate galaxies with AGNs using models for the probability of a galaxy hosting an AGN as a function of specific accretion rate. Our model is based on measurements indicating that the specific accretion rate distribution is a universal function across a wide range of host stellar mass with slope $\gamma_1\approx-0.65$ and an overall normalization that evolves with redshift. We test several simple assumptions to extend this model to high specific accretion rates (beyond the measurements) and compare the predictions for the XLF with the observed data. We find good agreement with a model that allows for a break in the specific accretion rate distribution at a point corresponding to the Eddington limit, a steep power-law tail to super-Eddington ratios with slope $\gamma_2=-2.1^{+0.3}_{-0.5}$, and a scatter of 0.38 dex in the scaling between black hole and host stellar mass. Our results show that samples of low luminosity AGNs are dominated by moderately massive galaxies ($\mathcal{M_*}\sim10^{10-11}\mathcal{M}_\odot$) growing with a wide range of accretion rates due to the shape of the galaxy stellar mass function rather than a preference for AGN activity at a particular stellar mass. Luminous AGNs may be a severely skewed population with elevated black hole masses relative to their host galaxies and in rare phases of rapid accretion.
\end{abstract}

\keywords{
galaxies: active -- galaxies: evolution -- X-rays: galaxies
}

\maketitle

 \section{Introduction}
 
The universe has evolved rapidly over the last $\sim 8$ billion years, since a redshift $z\sim1$.
The total star formation rate density, the rate at which new stars are being formed, has dropped by a factor 
 $\gtrsim10$ \citep[e.g.][]{Wilkins08,Zhu09,Rujopakarn10}.
A similar decline is seen in the total rate of accretion onto supermassive black holes (SMBHs), which is tracked by 
the luminosity density of Active Galactic Nuclei (AGNs) \citep[e.g.][]{Barger05,Silverman08,Aird10}. 
This correlation suggests that the processes regulating the formation of stars throughout galaxies
are somehow related to the processes that 
drive gas into their very central regions, feeding the growth of SMBHs and prompting periods of AGN activity.
It is thus vital to understand how the rapid decline of SMBH accretion since $z\sim1$ is related to the co-evolving properties of their host galaxies.

The AGN luminosity function provides the principal tracer of the distribution of SMBH accretion over the history of the universe.
A variety of wavebands and identification techniques have been used to measure the luminosity function of AGNs out to high redshifts ($z\sim6$)
\citep[e.g.][]{Boyle87,Page97,Richards06,Assef11,Ueda03,Ebrero09,Aird10}.
These studies have revealed a ``downsizing" behavior in the evolution of AGNs \citep[e.g.][]{Barger05}.
This downsizing is characterized by a rapid decline in the number density of the most luminous AGNs since a peak at $z\sim2$, while the number density of lower luminosity AGNs evolves much more weakly and peaks at lower $z$. 

To understand the underlying physical processes that drive the observed evolution of the AGN luminosity function it is vital to connect the AGN population to the galaxies that host them. 
A number of studies have indicated that AGNs are preferentially found in 
the most massive galaxies \citep[e.g.][]{Kauffmann03,Dunlop03,Schawinski07,Nandra07,Coil09,Hickox09,Georgakakis11b,Bongiorno12}.
These galaxies tend to have red optical colors and very low levels of current star formation.
Such trends suggest an association between the presence of an AGN and the quenching of star formation throughout a galaxy, possibly due to feedback from the AGN. 
A number of more recent studies, however, have shown that AGN hosts have a similar distribution of colors to galaxies of equivalent stellar masses \citep[e.g.][]{Silverman09b,Xue10,Cardamone10,Rosario13}.

\citet{Aird12} recently showed that the predominance of AGNs in more massive galaxies is in fact a selection effect;
the probability of a galaxy hosting an AGN is determined by a power-law distribution of specific accretion rates---the rate of accretion scaled by the stellar mass of the host---where the distribution itself does \emph{not} depend on stellar mass. 
AGNs are present in galaxies with a wide range of stellar masses, but those in more massive galaxies are simply more luminous.
\citet{Hainline12} and \citet{Bongiorno12} have presented similar findings, extending the investigations to higher redshifts.
Nevertheless, these results indicate that the overall distribution of the stellar masses of galaxies, traced by their stellar mass function, is a vital component in determining the distribution of observed AGN luminosities, traced by their luminosity function.

In this paper we develop a simple, observationally motivated model that connects the evolution of AGNs and their host galaxies at $0.2<z<1.0$. 
We start with the observed stellar mass function of galaxies and populate the galaxies with AGNs using a model for the probability of hosting an AGN as a function of specific accretion rate. 
Our model is based on measurements from \citet{Aird12} at low-to-moderate accretion rates. 
We extend the model to higher accretion rates using a number of well-motivated assumptions.
We thus predict the overall AGN luminosity function, which we then compare with the observed X-ray luminosity function of AGNs.
This comparison allows us to test and refine our model, providing important insights into the nature of the AGN population and the physical underpinnings of the AGN luminosity function.
We describe our observational datasets in Section \ref{sec:data} and use these to construct our model in Section \ref{sec:model}. 
Section \ref{sec:discuss} discusses our results and we state our overall conclusions in Section \ref{sec:conclusions}.

\begin{figure*}[t]
\plotone{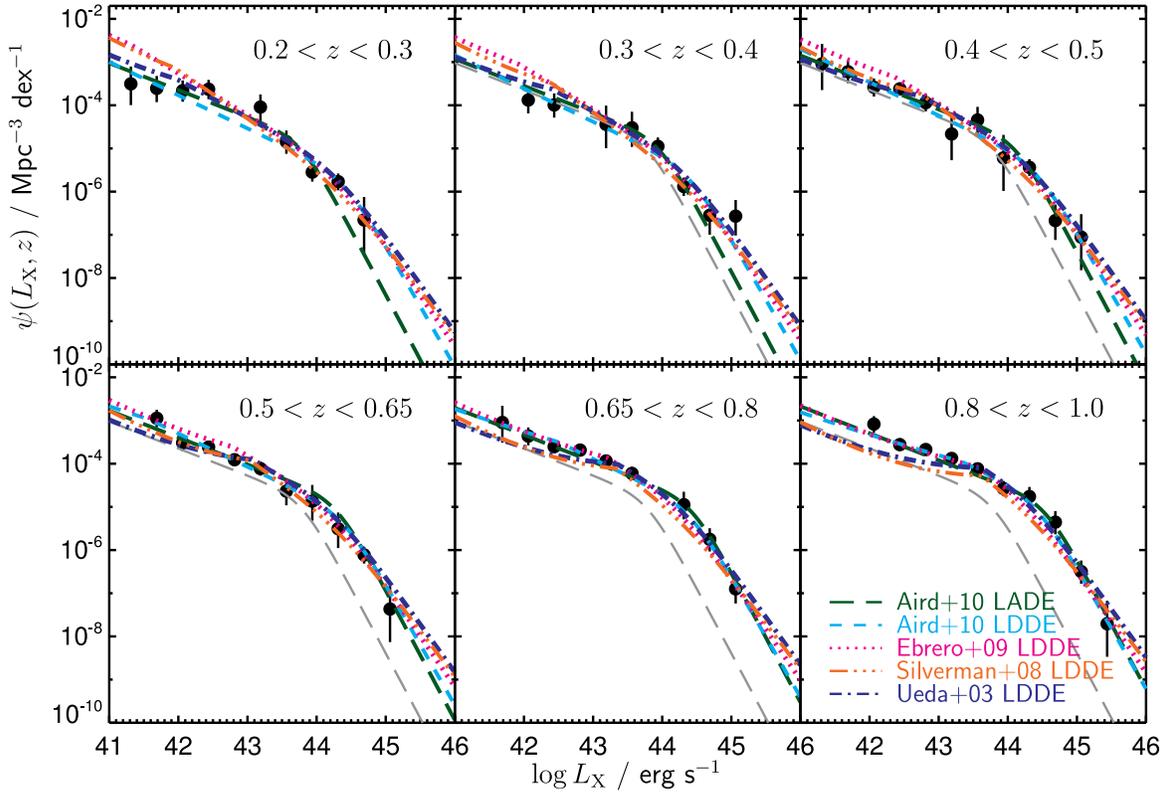}
\caption{
Comparison of the binned measurements of the hard (2--10 keV) X-ray luminosity function (XLF) from \citet{Aird10} (black circles, recalculated for the redshift bins shown here) and parametric model fits from various studies.
The grey dashed line indicates the \citet{Aird10} LADE model evaluated at $z=0.25$ (the center of the first redshift bin) and is replicated in every panel.
We also show the best fit Luminosity-Dependent Density Evolution (LDDE) parameterisations of the XLF from \citet[cyan]{Aird10}, \citet[magenta]{Ebrero09}, \citet[orange]{Silverman08} and \citet[dark blue]{Ueda03}.
While the details of the evolutionary behavior differ between the various parameterisations, all are approximately consistent with the observed data (black points). The evolution of the XLF from $z\approx 1$ to $z \approx 0.2$ is predominantly driven by a shift of the XLF to lower luminosities with decreasing redshift (i.e. luminosity evolution).
We adopt either the observed data points and their uncertainties (where available) or the range of the 5 different parameterisations to illustrate current measurements of the XLF and the uncertainties in subsequent plots.
}
\label{fig:xlf}
\end{figure*}

\section{Observational Datasets}
\label{sec:data}

\subsection{The galaxy stellar mass function}

To track the overall galaxy population we adopt recent measurements of the galaxy stellar mass function (SMF) from \citet{Moustakas13}.
The SMF was measured using the PRIsm MUlti-object Survey (PRIMUS), a 9.1 deg$^2$ low-resolution spectroscopic redshift survey of galaxies to $i\sim23$ \citep{Coil11,Cool13}.
From the parent sample of $\sim120,000$ robust redshifts, 
\citet{Moustakas13} constructed a flux-limited sample of $\sim40,000$ galaxies at $0.2<z<1.0$ over an area coverage of $5.5$ deg$^2$. 
Stellar masses were determined using fits to the observed spectral energy distributions (SEDs) based on ultra-violet, optical, near-infrared and mid-infrared photometry from \textit{GALEX}, \textit{Spitzer}/IRAC and a range of ground-based imaging campaigns \citep[see ][for full details]{Moustakas13}.
The SMF was constructed using the $V_\mathrm{max}$ method, carefully accounting for the PRIMUS targeting weights and redshift success rates, in six redshift bins between $z=0.2$ and $z=1.0$ (corresponding to roughly equal intervals of cosmic time).
For this study we adopt Schechter fits to the total SMF in each individual redshift bin, fully propagating the covariant uncertainties in the model fit parameters. 
A single Schechter form may not provide a full description of the shape of the SMF, but it does provide a good fit for the range of redshifts and stellar masses considered here.
The Schechter fits allow us to smooth between the $V_\mathrm{max}$ measurements in different stellar mass bins and perform moderate extrapolation to lower and higher stellar masses.

The \citet{Moustakas13} results indicate remarkably little evolution in the global SMF of galaxies in this redshift range:
the cumulative number density of $\mstel>10^{10}\msun$ galaxies has increased by $\sim30$\% since $z\sim0.6$ and the cumulative number density of $\mstel>10^{11}\msun$ galaxies has changed by $<\pm10$\% since $z\sim1$.
Much stronger, differing evolution is found when the sample is divided into quiescent and star-forming galaxies, placing strong constraints on the rates of star formation quenching that builds up the quiescent population. 
In this paper, however, we consider only the global SMF and attempt to reconcile the relative lack of evolution in the SMF of galaxies with the strong evolution observed in the AGN population. 

\subsection{The distribution of specific accretion rates}
\label{sec:accdist}

To populate galaxies with AGNs we adopt measurements of the distribution of SMBH accretion rates for a given stellar mass and redshift from \citet{Aird12}. 
In \citet{Aird12} we identified AGNs within the PRIMUS galaxy sample using hard (2--10 keV) X-ray observations of three of the PRIMUS fields.
We then studied the probability of a galaxy hosting an AGN as a function of the stellar mass.

We used hard X-ray selection to identify AGNs because X-ray emission from a moderately luminous AGN ($L_\mathrm{2-10keV}\gtrsim10^{42}$ \ergs) will dominate over other processes within the host galaxy (such as the combined emission from stellar X-ray binaries).
In addition, hard X-rays ($\gtrsim2$ keV) are able to penetrate moderate column densities of obscuring material at these redshifts (equivalent hydrogen column densities $\mathrm{N_H}\lesssim10^{23}$ cm$^{-2}$). Thus, hard X-ray selection provides a reasonably accurate probe of the AGN luminosity and should track the bulk of the AGN population.
The effect of the (highly variable) flux limits of the X-ray observations over the three PRIMUS fields was fully accounted for by the \citet{Aird12} work.
We note that hard X-ray selection will fail to identify the most heavily obscured, Compton-thick sources. 
It remains unclear how large a fraction of the population is hidden by such high levels of obscuration.
However, our measurements of the overall X-ray luminosity function (which we compare with our model predictions, see Section \ref{sec:xlf} below) also use hard X-ray selection to identify AGN and thus are subject to the same incompleteness. 

\citet{Aird12} showed that the probability of a galaxy with a given stellar mass and redshift hosting an AGN could be described by a single power law distribution of specific accretion rates. This distribution is independent of stellar mass and evolves strongly with redshift, 
\begin{equation}
p(\lambda \giv \mstel,z) \dd \log\lambda = A \lambda^\gamma \left(\frac{1+z}{1+z_0}\right)^\beta \dd\log\lambda
\label{eq:pldist}
\end{equation}
where $\lambda\propto L_\mathrm{bol}/\mstel$ denotes the \textit{specific accretion rate}, the rate at which mass is accreted by the SMBH (traced by the AGN bolometric luminosity, $L_\mathrm{bol}$) scaled by the total stellar mass of the host galaxy, \Mstel, and 
$p(\lambda \giv \mstel,z)$ is the probability density per logarithmic interval in $\lambda$ of a galaxy at a given redshift hosting an AGN with specific accretion rate, $\lambda$.
We measured the slope of the probability distribution function as $\gamma=-0.65\pm0.04$, a strong redshift dependence ($\beta=3.5\pm0.5$) 
\amend{
and a normalization factor $\log_{10} A=-3.15\pm0.08$}. Our work spanned a wide range of stellar masses ($9.5<\log \mstel/\msun<12$) and focussed on low-to-moderate luminosity AGNs ($42<\log \lx \;/\;\mathrm{erg\;s^{-1}}<44$). 
Thus our results spanned a limited range in specific accretion rate, $-3\lesssim\log \lambda\lesssim-1$, where $\lambda$ is defined relative to the Eddington limit assuming a single scaling between SMBH mass and the total stellar mass of the host galaxy 
\amend{\citep[$\mbh\approx 0.002\mstel$, based on][where we also assume $\mstel \approx \mathcal{M}_\mathrm{bulge}$]{Marconi03}},
\begin{equation}
\lambda=       \dfrac{L_\mathrm{bol}}{L_\mathrm{Edd}} 
= \frac{L_\mathrm{bol}}{ 1.3\; \times \; 10^{38} \; \mathrm{erg\; s}^{-1} \times 0.002 \dfrac{\mstel}{\msun}}.
\label{eq:ledd}
\end{equation}
These findings form the basis of our study in this paper.
We investigate different models (see Section \ref{sec:model} below) to extend the $\lambda$ distribution to higher accretion rates and compare with the AGN luminosity function over a wide luminosity range.

Broad-line QSOs were identified using the PRIMUS spectra and excluded from the \citet{Aird12} samples. 
In these sources the AGN emission dominates over the host galaxy at optical wavelengths, precluding measurements of the stellar mass of the host galaxy.
Over the X-ray luminosities probed by \citet{Aird12}, $\lx=10^{42-44}$ \ergs, the broad-line QSOs correspond to $\lesssim 20$\% of the AGN population. 
However, the broad-line fraction increases with luminosity \citep[e.g.][]{Steffen03,Treister09b} and 
thus excluding broad-line QSOs may bias the slope of the $\lambda$ distribution measured by \citet{Aird12}.
On the other hand, previous studies have found that QSOs may be predominantly high accretion rate sources, accreting close to their Eddington limit (e.g. \citealt{Kollmeier06,Trump11}, but see also \citealt{Kelly13}) and dominate the population at $\lx>10^{44}$ \ergs. Our models (described below) that extrapolate the $\lambda$ distribution to higher accretion rates may therefore account for the bulk of the broad-line QSO population.
We discuss remaining issues regarding the exclusion of broad-line QSOs in Section \ref{sec:discuss}.

\subsection{The X-ray luminosity function of AGNs}
\label{sec:xlf}

The AGN luminosity function traces the overall distribution of accretion activity in AGNs in terms of the observed luminosities at a particular waveband. 
Any model that populates galaxies with accreting SMBHs must be consistent with this observational constraint on the total population. 
In this paper we compare our model to measurements of the 2--10 keV X-ray luminosity function (XLF) of AGNs at $0.2<z<1.0$.

In Figure \ref{fig:xlf} we show binned estimates of the XLF from \citet{Aird10}, recalculated in the redshift bins used for this study (black points with error bars). 
This work applied a sophisticated Bayesian methodology to account for uncertainties in photometric redshifts and X-ray flux estimates, track the X-ray sensitivity, correct for Eddington bias, and correct for incomplete optical identifications.
The binned estimates shown in Figure \ref{fig:xlf} are calculated using the $N_\mathrm{obs}/N_\mathrm{mdl}$ method \citep{Miyaji01}, which corrects for the effects of the X-ray sensitivity over the width of the bin. 
Error bars represent $1\sigma$ equivalent Poissonian errors for the number of objects in a bin.
We also show a number of best-fit parametric models, including both the luminosity-dependent density evolution (LDDE) model determined by \citet{Aird10} (cyan dashed line) and the conceptually simpler Luminosity And Density Evolution (LADE) model where the XLF evolves in both luminosity and density but retains the same shape at all redshifts (green  line). 
\citet{Aird10} concluded that it was not possible to statistically distinguish between these models and thus advocated the simpler LADE model. 
In addition, we show the best fit LDDE parameterizations from \citet{Ueda03}, \citet{Ebrero09}, and \citet{Silverman08}.

All of the model parameterizations demonstrate the same basic evolutionary trend,
namely a reduction in the number density of more luminous AGNs, which can be characterized by an overall shift of the AGN population to lower luminosities as cosmic time progresses. 
This luminosity evolution dominates the evolution described by the LADE parameterization of \citet{Aird10} at these redshifts ($z<1$) and has also been quantified by pure luminosity evolution models in earlier works \citep[e.g.][]{Barger05}.
However, the LDDE parameterizations indicate that the shape of the XLF may be changing with redshift.
Given these differences in the parameterizations of the XLF, we choose to adopt the observed, binned data points and their uncertainties when comparing with our model predictions. 
Outside the range of the \citet{Aird10} data ($\lx\gtrsim10^{45}$\ergs at these redshifts) we conservatively take the minimum and maximum of the 5 model parameterizations shown in Figure \ref{fig:xlf} to represent current uncertainties in the XLF. 
This XLF estimate is shown by the light-green hatched region in subsequent plots.

\begin{figure*}
\plotone{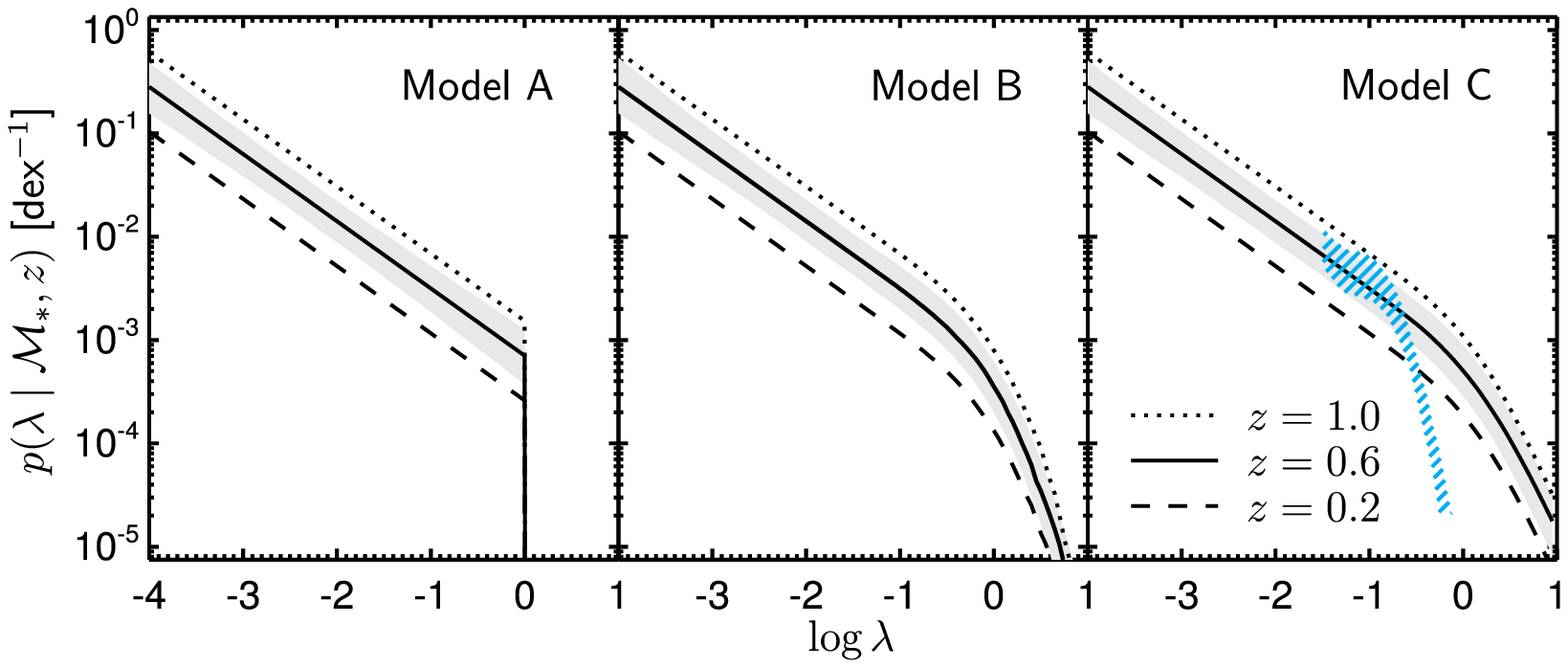}
\caption{Assumed distributions of specific accretion rates (for a sample of galaxies at a given stellar mass and redshift), based on measurements (at $-3\lesssim \log\lambda\lesssim -1$) by \citet{Aird12}. 
The units assume $\lambda={L_\mathrm{bol}}\left({1.3 \times 10^{38} \mathrm{erg \; s^{-1}} \times 0.002 \frac{\mstel}{\msun}}\right)^{-1}$.
In model A we assume a single scaling between the SMBH mass and stellar mass of the host galaxy; 
thus, $\lambda$ directly translates into the Eddington ratio in the assumed units and we apply a strict cutoff at the Eddington limit, $\log \lambda=0$. 
The lines indicate the distribution at $z=1.0$ (dotted), $z=0.6$ (solid) and $z=0.2$ (dashed) showing the evolution in the probability of a galaxy hosting an AGN with redshift as measured by \citet{Aird12}. 
The grey region indicates the uncertainty in the model fit to the \citet{Aird12} measurements, including an additional $\pm$50\% uncertainty in the normalization.
In model B we allow for an intrinsic lognormal scatter of 0.38 dex in the scaling between SMBH mass and total host stellar mass.
This scatter softens the cutoff in terms of specific accretion rate and predicts a population of SMBHs with high masses relative to their host accreting close to their Eddington limit, which thus have high specific accretion rates.
In model C we also allow for a steep power-law distribution to higher specific accretion rates, corresponding to a rare population of SMBHs accreting at rates above their Eddington limit.
\moreamend{The blue hashed region in this final panel shows the conditional probability distribution of Eddington ratios for QSOs in the SDSS at $z=0.6$ and $\mbh=5\times 10^8 \msun$ from \citet{Kelly13}, rescaled to match the normalization of model C at $\log \lambda=-1$. 
This distribution has a similar slope at low $\lambda$ to model C, but has a break below the Eddington limit ($\log \lambda\approx-0.5$) with a rapid decline to higher accretion rates. These differences are discussed in Section \ref{sec:nature}.}
}
\label{fig:accdists}
\end{figure*}

\section{An observationally motivated model to connect the AGN and galaxy populations}
\label{sec:model}

In this section we present a simple, observationally motivated model to populate galaxies with AGNs.
We start with our measurements of the galaxy SMF and convolve with a model for the probability of hosting an AGN as a function of specific accretion rate.
This convolution provides us with a prediction for the XLF of the AGN population, which we can compare with the observations.
Our model for the XLF can be expressed as
\begin{eqnarray}
\psi(L_\mathrm{X}, z) & =  & \phi(\mstel , z) * p(\lambda \giv \mstel,z) \\
				   & = & \int \phi(\mstel , z) \; p\big( \lambda(L_\mathrm{X}, \mstel) \giv \mstel,z\big) \dd \log \mstel \nonumber
\end{eqnarray}
where $\phi(\mstel, z)$ is the galaxy SMF and $p(\lambda \giv \mstel,z)$ describes the probability density per logarithmic $\lambda$ interval for a galaxy of given \Mstel\ and $z$ hosting an AGN with a specific accretion rate, $\lambda$.\footnote{
\amend{We note that the probability distribution defined by equation \ref{eq:pldist} diverges at low $\lambda$. We therefore set a minimum specific accretion rate for a given redshift, $\lambda_\mathrm{min}(z)$, such that
$\int_{\lambda_\mathrm{min}(z)}^{\infty} p(\lambda \giv \mstel,z) \dd \log\lambda = 1.0$. The position of this cutoff generally falls outside of the range covered by our datasets so has no practical effect on the results of this paper.}
}
This equation allows us to predict the XLF, $\psi(L_\mathrm{X}, z)$,
assuming the luminosity-dependent bolometric corrections of \citet{Hopkins07b} to convert from $\lambda$ to an X-ray luminosity, $L_\mathrm{X}$, for a given \Mstel.
In Figure \ref{fig:accdists} we show our three models for $p(\lambda \giv \mstel,z)$, all of which are based on the power-law distribution measured by \citet{Aird12} that is independent of stellar mass but evolves strongly with redshift (see Equation \ref{eq:pldist} and Section \ref{sec:accdist} above).

\subsection{Model A}
\begin{figure*}
\plotone{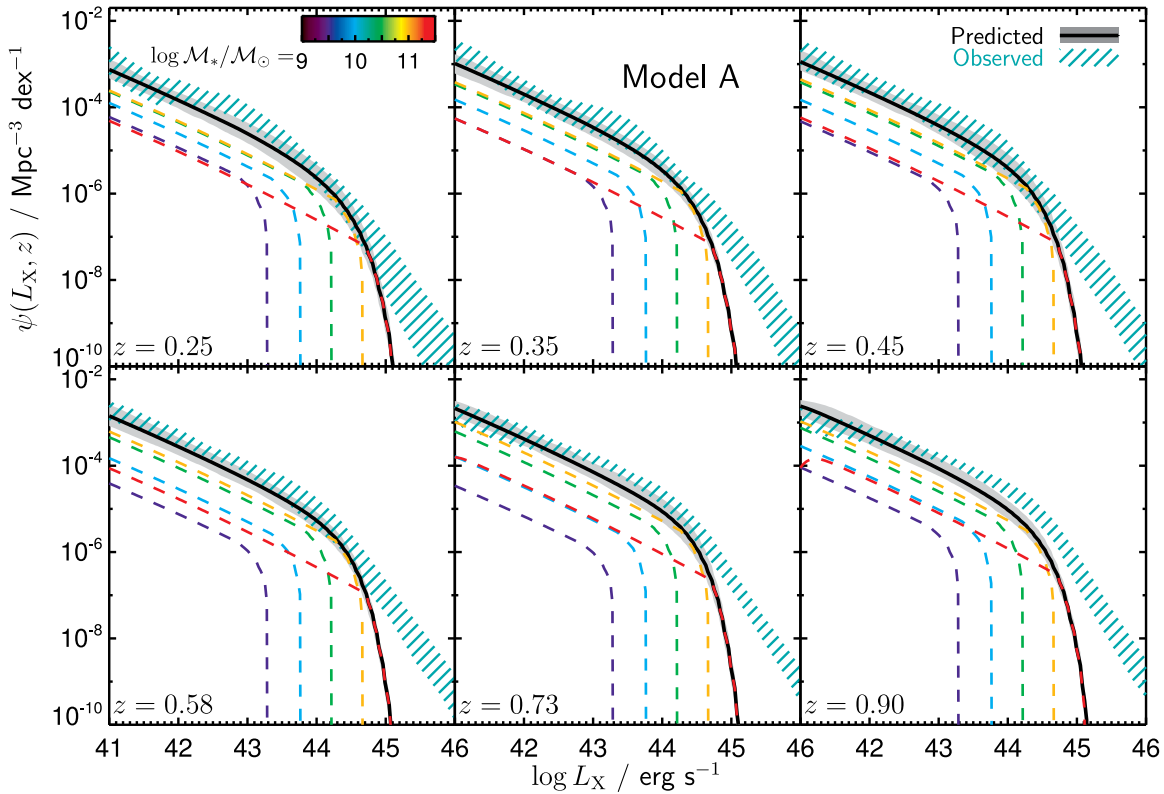}
\caption{Predictions for the XLF (black solid line, grey region shows propagated $1\sigma$ uncertainty in the model) at a range of redshifts based on the galaxy SMF and assuming model A for the distribution of accretion rates. Model A assumes the mass of the SMBH has a single scaling with the galaxy stellar mass and thus the distribution of specific accretion rates has a power-law form with a sharp cutoff corresponding to the Eddington limit. 
The colored lines show the contributions to the XLF from galaxies over limited ranges of stellar mass.
We compare to direct observations of the XLF (green hatched regions, which track the uncertainty in the observations -- see Figure \ref{fig:xlf} and Section \ref{sec:xlf} for details).
Our model successfully predicts the faint end of the XLF ($L_\mathrm{X}\lesssim 10^{44}$ \ergs) but falls over much more rapidly than the observed slope for more luminous AGNs and fails to reproduce the evolution in the break of the XLF.
} 
\label{fig:xlf_modA}
\end{figure*}

We begin with a simple model (A), where we assume a single scaling applies between SMBH mass and the total stellar mass of the host galaxy at all redshifts.
We take the scaling between SMBH mass and galaxy mass as $\mbh\approx 0.002\mstel$ from \citet{Marconi03}, where we have further assumed that $\mstel \approx \mathcal{M}_\mathrm{bulge}$.
Thus, we can directly translate the specific accretion rate into an Eddington ratio for the SMBH (see Equation \ref{eq:ledd}).
With this scaling assumed, we apply a strict cut to our specific accretion rate distribution at the Eddington limit, $\log \lambda=0$, as shown in Figure \ref{fig:accdists} (left panel). 
The grey region tracks the uncertainty in our specific accretion rate distribution.
We estimate the uncertainties using the errors in the best-fit model parameters from \citet{Aird12}. 
However, these errors assume a fixed model form and do not fully account for the observational uncertainties in 
the measurements of the $\lambda$ distribution at a given stellar mass or redshift. 
We therefore increase the uncertainty in the overall normalization by $\sim \pm 0.2$ dex, which provides better agreement between our model uncertainty and the errors in the individual binned measurements of $p(\lambda \giv \mstel,z)$ in figure 7 of \citet{Aird12}.
This increase also allows for additional systematic uncertainties in our model, such as the possible contribution of unobscured sources. 
In Figure \ref{fig:xlf_modA} we compare the prediction assuming our model A for the specific accretion rate distribution (thick black line with the solid grey region indicating the uncertainty) with the observational constraints on the XLF (green hatched region).
Our simple, observationally motivated model is able to successfully reproduce the faint-end of the XLF at all redshifts to $z\sim0.7$. 
The predominant evolution with redshift is accounted for by the strong evolution in the normalization of $p(\lambda \giv \mstel,z)$,
which produces a change in the normalization of our predicted XLF. 
However, we fail to reproduce the correct behavior at the bright end of the XLF ($L_\mathrm{X}\gtrsim10^{44.5}$ \ergs). 
Our model predicts an exponential decline in the XLF at high luminosities, a consequence of the Schechter form of the galaxy SMF and the sharp cutoff we apply in the specific accretion rate distribution. 
We note that our model does not reproduce any of the (relatively minor) changes that may be observed in the evolution of the faint-end slope of the XLF with redshift nor any change in the break of the XLF.
At $z\gtrsim0.7$ we underpredict the number density of $L_\mathrm{X}\approx10^{43.5-44.5}$\ergs\ AGNs, and the observed faint-end slope is somewhat flatter than our prediction.

\begin{figure*}
\plotone{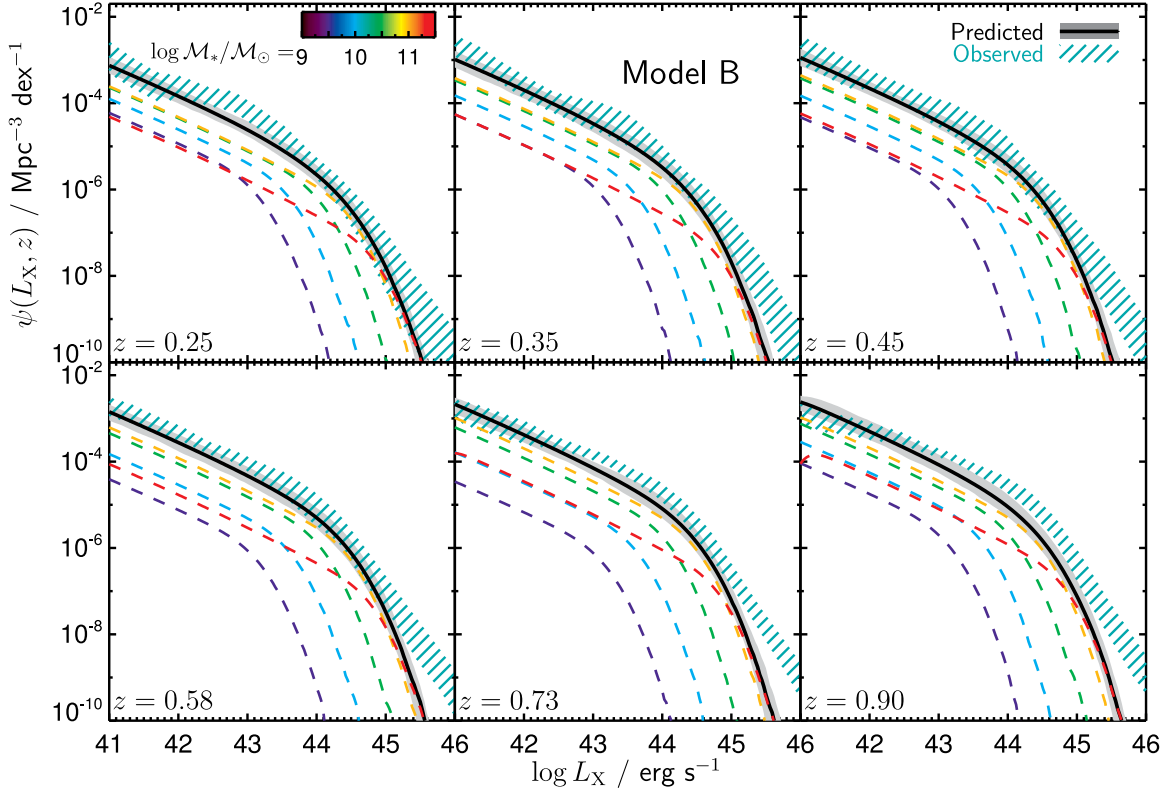}
\caption{As Figure \ref{fig:xlf_modA} but assuming model B for the distribution of specific accretion rates.
We account for intrinsic scatter of $0.38$ dex in the $\mbh-\mstel$ relation to calculate the Eddington limit, where we truncate the probability of a galaxy hosting an AGN.
This scatter softens the cut off in the distribution of specific accretion rates and flattens the predicted bright-end slope of the XLF.
Our prediction still falls off more rapidly than the observed XLF at high $L_\mathrm{X}$. 
}
\label{fig:xlf_modB}
\end{figure*}

\begin{figure*}
\plotone{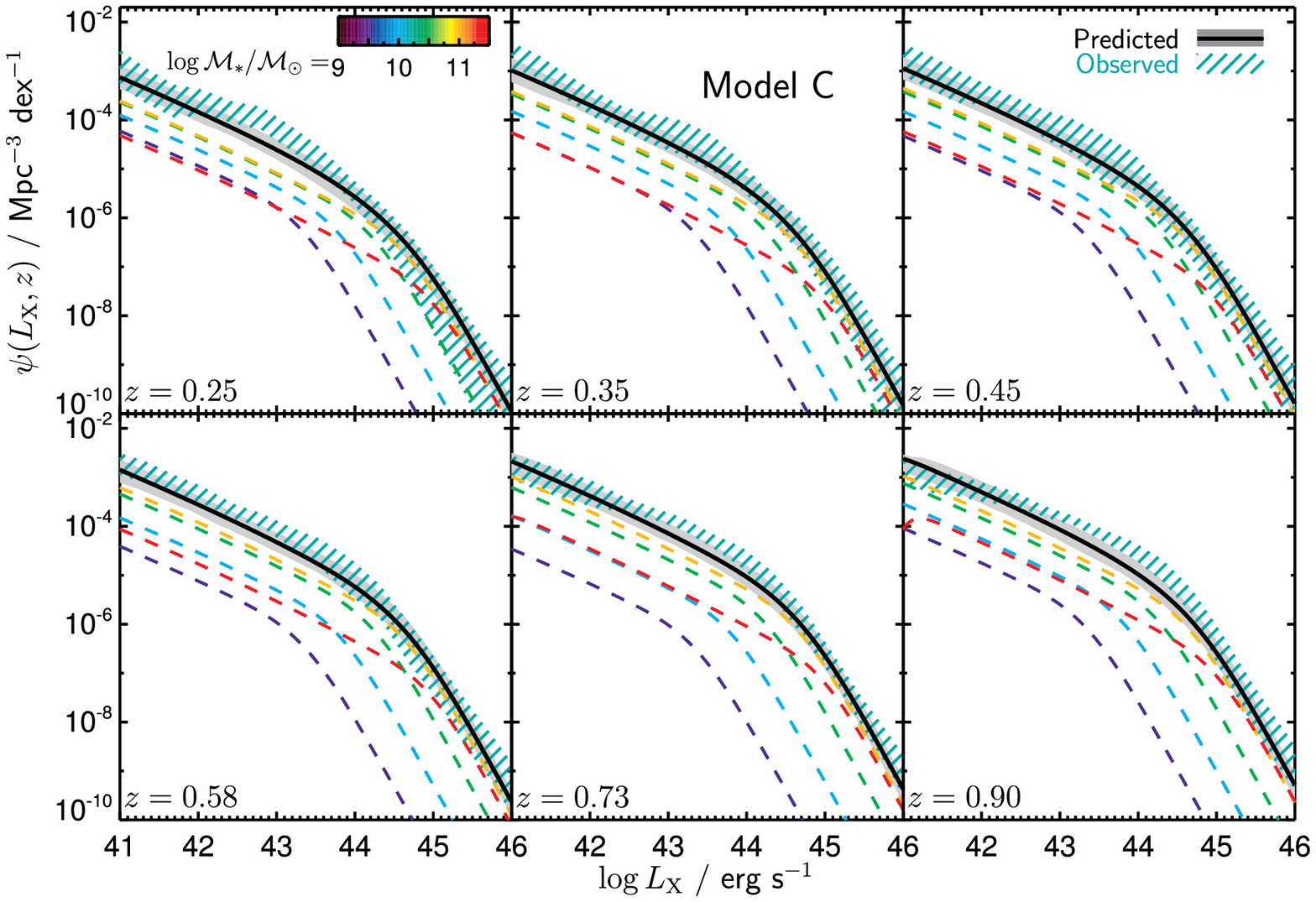}
\caption{As Figure \ref{fig:xlf_modB}, but assuming model C for the distribution of specific accretion rates.
In addition to the scatter in the $\mbh-\mstel$ relation, in this model we also include a steep power law tail to high values of $\lambda$, corresponding to super-Eddington accretion rates.
Our model is able to reproduce the dominant features of the observed XLF  from $z\sim1$ to $z\sim0.2$.
}
\label{fig:xlf_modC}
\end{figure*}

\subsection{Model B}
\label{sec:modelB}
In our second model (B), we adopt a more realistic scaling relation between the SMBH mass, \Mbh, and host stellar mass, \Mstel, 
that allows for intrinsic scatter in the relationship.
We adopt an intrinsic scatter of 0.38 dex, consistent with measurements of the local relationship between spheroid stellar mass, $\mathcal{M}_\mathrm{spheroid}$, and \Mbh\ by \citet{Bennert11b}.
This work used updated dynamical \Mbh\ estimates from \citet{Gultekin09} for a sample of 18 local elliptical and S0 galaxies;
spheroid stellar masses were estimated using the $J$, $H$ and $K$ magnitudes measured by \citet{Marconi03} and the \citet{Auger09} Bayesian stellar mass estimation code.
We note that the \citet{Bennert11b} relationship was measured using the spheroid (bulge) stellar masses, but we assume that the same intrinsic scatter can be applied to the total stellar mass - SMBH mass relationship.

In model B we assume this 0.38 dex scatter in the \Mbh-\Mstel\ relationship to convert between $\lambda$ and Eddington ratio, applying a sharp cutoff in the probability of a galaxy hosting an AGN above the Eddington limit.
The scatter in \Mbh-\Mstel\ results in a much softer cutoff in terms of specific accretion rate, $\lambda$ (see Figure \ref{fig:accdists}, center). 
The tail in $p(\lambda \giv \mstel,z)$ at high specific accretion rates ($\log \lambda>0$) can be attributed to the probability of finding galaxies with elevated SMBH masses relative to the host (due to the intrinsic scatter) accreting close to their Eddington limits.

Figure \ref{fig:xlf_modB} compares the observed XLF and our prediction using model B.
The softer cutoff in the $\lambda$ distribution due to the scatter of the \Mbh--\Mstel\ relationship results in a smoother break in our predicted XLF compared to model A (Figure \ref{fig:xlf_modA}). 
Furthermore, at the bright end of the XLF ($\lx \gtrsim10^{44.5}$ \ergs), AGNs with typical mass SMBHs are outnumbered by the small fraction of lower stellar mass galaxies with elevated SMBH masses, due to the intrinsic scatter in the \Mbh-\Mstel\ relation and the steep slope of the SMF at high stellar masses  \citep[$\mstel\gtrsim10^{11}\msun$, see also ][]{Somerville09}.
This effect leads to a somewhat flatter predicted bright-end slope than found with Model A.
Nonetheless, our prediction still declines more rapidly than the observed XLF at $L_\mathrm{X}\gtrsim 10^{45}$\ergs, especially at high redshifts ($z\gtrsim0.5$).

\subsection{Model C}

Prompted by the lack of agreement at the bright end of the XLF using model B, our third model (C) removes the requirement of a sharp cutoff at the Eddington limit and instead adopts a steep power-law distribution with a tail to very high Eddington ratios, while retaining the 0.38 dex scatter in the \Mbh--\Mstel\ scaling.
The Eddington limit only strictly applies to a spherical geometry and studies of accretion disk physics do allow for scenarios when the Eddington limit is violated \citep[e.g.][]{Jaroszynski80,Abramowicz88,Begelman06}.
Indeed, SMBHs accreting above their Eddington limits are rare but are found in sufficiently large samples of AGNs \citep[e.g.][]{Mclure04,Kelly13}.
The slope of this high accretion rate tail has not been directly measured in our observed specific accretion rate distribution \citep[although hints of such a turnover may be seen by ][]{Bongiorno12}. 
We therefore introduce this as a single free parameter, $\gamma_2$, in our model that we fit using the original binned data points in our observed XLF (shown in Figure \ref{fig:xlf}). 
We constrain the slope to be $\gamma_2=-2.1^{+0.3}_{-0.5}$. 
\amend{
We note that while the Eddington limit does provide a physically motivated scaling for our distributions, a turn over at precisely this point may not be realistic. If we allow the break in the distribution to be an additional free parameter in our model C, which we constrain by fitting the observed XLF, we find it is $\log \lambda_\mathrm{break}=0.2\pm0.2$, consistent with $\log \lambda_\mathrm{break}=0.0$.
A break at around the Eddington limit  is clearly needed to reproduce the bright end of the observed XLF (given the other assumptions and restrictions of our model); a break in the distribution at lower accretion rates would not produce sufficient luminous AGNs.
We thus retain a one-parameter model with a fixed break at $\log \lambda_\mathrm{break}=0.0$.}

\amend{
We also note that the shape of $p(\lambda \giv \mstel,z)$ at high accretion rates ($\log \lambda \gtrsim 0.1$) may be sensitive to the assumed form of the bolometric correction.
Recent work has indicated that X-ray bolometric corrections may be primarily a function of the Eddington ratio itself \citep[e.g.][]{Vasudevan07,Lusso12,Jin12}, rather than the AGN luminosity as we have assumed \citep[following ][]{Hopkins07b,Marconi04}.
The bolometric correction increases with increasing Eddington ratio and may be $\gtrsim 100$ at the highest accretion rates \citep[$\lambda>1$,][]{Jin12}. 
A high bolometric correction means a low fraction of the AGN light is emitted at X-ray wavelengths.
Thus, recovering the luminous AGN population may actually require a steeper slope for the super-Eddington tail of the accretion rate distribution; an even higher accretion rate is required to produce the observed X-ray luminosities.
However, the slope of the super-Eddington tail of the accretion rate distribution remains poorly constrained due to the limited samples of luminous, hard X-ray selected AGNs that enter our XLF, and thus the exact form of the bolometric correction does not have a substantial impact on our findings.
}

In Figure \ref{fig:xlf_modC} we compare our prediction from model C with the observed XLF. 
With model C we are able to produce an XLF that is in good agreement with the observed data (considering the uncertainties in both our prediction and the observations). 
However, there are minor discrepancies between our observations and our model for the XLF.
Our model underpredicts the number density of $L_\mathrm{X}\approx 10^{43.5-44.5}$\ergs\ AGNs at $z\gtrsim0.7$, although the discrepancy is only at the $\sim 2\sigma$ level when considering the uncertainties in both the observed XLF and our model prediction (propagated from the SMF and measurements of the specific accretion rate distribution). 
In addition, the evolutionary behavior of our model is predominantly a density evolution. 
We do not reproduce the luminosity-dependent evolution found in a number of studies of the XLF, 
although our model prediction does provide reasonable agreement with the observed XLF considering the current uncertainties. 
We discuss these issues further in Section \ref{sec:evol}.

Our observationally motivated model allows us to interpret the observed AGN population at $0.2<z<1.0$, as traced by the XLF.
As shown by the colored lines in Figure \ref{fig:xlf_modC}, our model predicts that the low $L_\mathrm{X}$ population is dominated by AGNs in hosts with stellar masses $\sim10^{10.5-11}$ \Msun\ accreting over a wide range of specific accretion rates \citep[consistent with the predominance of such host galaxies in X-ray selected samples, e.g.][]{Xue10,Bongiorno12}.
Around the break in the XLF the population is dominated by moderately massive galaxies ($\mstel\sim 10^{11}\msun$) accreting close to their Eddington limit. 
The bright end of the XLF is dominated by massive galaxies ($\mstel\gtrsim10^{11}\msun$)
accreting close to or above the Eddington limit.
Assuming a 0.38 dex scatter in the \emph{intrinsic} $\mbh-\mstel$ relation, we predict that 80\% of detected AGN with $\lx>10^{44.5}$ \ergs\ will have SMBH masses that are above the fiducial $\mbh-\mstel$ relation; the predicted median $\mbh/\mstel$ for this luminous population is $\sim0.25$ dex higher than the median for the overall galaxy population.
This effect can be explained by the steep slope of the SMF at high masses and the intrinsic scatter in the \Mbh--\Mstel\ relation.
It does not indicate that SMBH masses are intrinsically higher in AGN host galaxies.
Finally, the evolution in the normalization of the XLF is explained by a rapid drop in the probability of a galaxy hosting an AGN at later times for all specific accretion rates and across all stellar masses.

\section{Discussion}
\label{sec:discuss}

We have shown how with a simple, observationally motivated model we are able to explain the predominant features of the XLF of AGNs and its evolution over $0.2<z<1.0$.
We start with the observed SMF of galaxies (and its very mild evolution) from PRIMUS \citep{Moustakas13}
and populate these galaxies with AGNs based on a model for the probability of hosting an AGN as a function of specific accretion rate.
Our model is based on the observations at low-to-moderate accretion rates (equivalent to Eddington ratios $\sim10^{-3}-0.1$) and over a wide range of stellar masses ($9.5<\log \mstel/\msun<12$) from \citet{Aird12}.
We extend our model to higher accretion rates (equivalent to Eddington ratios $\gtrsim0.1$) using a small number of straightforward assumptions.
Our final model has one free parameter---the slope of the distribution at super-Eddington accretion rates---which we fit using the observed
bright end of the XLF. 
These results provide new important insights into the nature of the AGN population and its evolution since $z\sim1$.

\subsection{The nature of the AGN population}
\label{sec:nature}

At low X-ray luminosities (the faint end of the XLF), the number density of AGNs is dominated by moderately massive galaxies ($\mstel \sim 10^{10.5-11} \msun$) with a range of specific accretion rates given by the power-law distribution with slope $\approx-0.65$ measured by \citet{Aird12}.
\amend{
The model predicts that X-ray flux limited samples (probing low X-ray luminosities at $z\lesssim1$) will identify AGNs with a distribution of host stellar masses that peaks at $\log \mstel/\msun \sim 10.5-11$, with the majority of sources in moderately massive galaxies ($10<\log \mstel/\msun<11.5$), consistent with observed samples \citep[e.g.][]{Alonso-Herrero08,Hickox09}.}
However, the predominant contribution by moderate mass galaxies is due to the shape of the SMF.
As shown in \citet{Aird12}, AGN activity does not appear to have a preference for a particular stellar mass range.
In our model, galaxies with lower or higher stellar masses are just as likely to host an AGN with the same distribution of accretion rates, but these sources do not constitute a significant fraction of the overall AGN number density.

The bright end of the XLF probes a different regime of the AGN population.
In our model C, we propose a break in the accretion rate distribution at a point corresponding to the Eddington limit and a steep power-law tail that allows for a rare population of sources with very high (super-Eddington) accretion rates.
We also allow for an intrinsic scatter in the scaling between the SMBH mass and the stellar mass of the host galaxy. 
We find that the bright end of the XLF consists mostly of massive galaxies ($\mstel\gtrsim10^{11}\msun$) accreting close to or above their Eddington limits.
This population is dominated by AGNs with SMBH masses that are elevated relative to the fiducial \Mbh--\Mstel\ relation, which is a result of scatter in the scaling relation and the steep decline of the SMF at high masses. 
\citet{Lauer07} discussed how such a bias could affect studies of SMBH mass scaling relations using broad line AGNs outside the local universe. 
In fact, recent studies have found tentative evidence for strong evolution in the scaling between \Mbh\ and host spheroid mass (e.g. \citealt{Treu07,Jahnke09,Merloni10,Bennert11}, \amend{although see also \citealt{Alexander08,Cisternas11}}). Our model, which assumes a constant scaling and scatter in the \Mbh-\Mstel\ relation, may provide a simple explanation for such findings, although a full investigation is beyond the scope of this paper 
\citep[\amend{see also}][]{Schulze11,Volonteri11, Salviander13}.

We note that in our study of specific accretion rates \citep[see Section \ref{sec:accdist}]{Aird12} we did not measure $p(\lambda \giv \mstel,z)$ for high accretion rate sources ($\log\lambda\gtrsim-1$).
In fact, our study excluded broad-line AGNs and was limited to moderate luminosity sources ($\lx=10^{42-44}$\ergs). 
We demonstrated in \citet{Aird12} that the exclusion of broad-line AGNs did not significantly bias our measurement of the slope of the specific accretion rate distribution at $-3\lesssim\log\lambda\lesssim-1$.
In fact, broad-line AGNs only constitute a significant fraction of the AGN population at high luminosities \citep[$\lx\gtrsim10^{44}$ \ergs, e.g.][]{Barger05,Trump09b}.
Measurements of $p(\lambda \giv \mstel,z)$ at $\log\lambda\gtrsim-1$, including broad-line and luminous ($\lx>10^{44}$ \ergs) AGN,
\amend{
are vital to confirm our model relating AGNs and their host galaxies, our extrapolation of the accretion rate distribution to the highest accretion rates, and our interpretation of the XLF. 
Such studies present a number of difficulties due to the relative rarity of such sources, the strong biases that can affect the properties of the observed population, and the challenge of disentangling the properties of luminous AGNs and their host galaxies.
}

\amend{
However, the high luminosity AGN population is probed by large area optical surveys, such as the SDSS \citep[e.g.][]{Mclure04,Ross12}, although such surveys only identify broad-line (Type 1) AGNs and will miss obscured, albeit intrinsically luminous, Type 2 sources.
Our model does not distinguish Type 1 AGNs, making direct comparisons between our model and measurements of the 
overall SMBH mass function or Eddington ratio distribution functions (ERDFs) of Type 1 AGNs difficult.
Nevertheless, we can compare the distributions of Eddington ratios for samples of Type 1 AGNs selected over limited luminosity ranges, such as those presented by \citet{Kollmeier06} or \citet{Shen08}, to predictions for the observed distributions based on our model.
Qualitatively, we find reasonable agreement for the distributions: our model predicts that luminous AGNs ($L_\mathrm{bol}\gtrsim10^{45.5}$ \ergs) selected in narrow bins of luminosity are predominantly at fairly high Eddington ratios ($\log \lambda\gtrsim-1.5$) with roughly lognormal distributions of widths $\sim0.3-0.5$ dex.
However, our model predicts mean accretion rates that are up to $\sim0.5$ dex higher than the observed distributions and a much larger fraction of AGNs accreting at super-Eddington rates -- as high as  50\% of the population for the most luminous AGNs ($L_\mathrm{bol}\gtrsim10^{46.5}$ \ergs). 
Such sources could correspond to the luminous Type 2 AGN population, indicating such sources are the most rapidly accreting sources \citep[e.g.][]{Warner04,Urrutia12}. 
However, the uncertainties in virial SMBH mass estimates, used to derive the Eddington ratios, can introduce substantial biases into the observed distributions \citep[e.g.][]{Shen08,Shen10,Schulze11}, potentially bringing them more in line with the 
predictions of our model. 
We also note that current uncertainties in the absolute scaling between SMBH mass and host mass could lead to systematic differences between our model predictions and these observations. 
Indeed, recent measurements of SMBH scaling relations \citep[e.g.][]{Graham12,Graham13} indicate that the scaling may not be constant with stellar mass and could be as high as $\mbh/\mstel\approx0.005$ for the highest mass objects,}
\moreamend{although some of these differences could be due to different normalizations in the scaling relations for different types of host galaxy \citep[e.g.][]{McConnell13}.}
\amend{
A larger \Mbh--\Mstel\ scaling for certain galaxies could move the break in the accretion rate distribution below the Eddington limit, which could allow better agreement between our results and these studies of optical Type 1 AGNs.
}

\amend{
Recent studies \citep[e.g.][]{Schulze10,Kelly13,Nobuta12} have presented measurements of the SMBH mass functions and ERDFs of Type 1 AGNs that model the completeness effects and biases of optical surveys. While such studies provide vital insight into the Type 1 AGN population, direct comparisons to our model are not possible as the Type 2 AGNs and low accretion rate sources dominate the space densities in most regimes. Nevertheless, such studies do find a SMBH mass function with a substantially flatter slope at high masses than the underlying galaxy SMF and a ERDF that continues to rise at low accretion rates with a slope $\sim-0.8$, in reasonable agreement with our model. 
A turn over in the ERDF and steep decline towards high accretion rates is also found, although in contrast to our model this comes in at $\lambda \gtrsim0.1$, substantially below the Eddington limit assumed in our model.}
\moreamend{This is illustrated in panel C of Figure \ref{fig:accdists} where we over-plot measurements of the distribution of Eddington ratios of Type 1 AGNs from the SDSS at fixed \Mbh\ from \citet{Kelly13}, rescaled to match the normalization of our model C at $\log \lambda=-1$.\footnote{
\moreamend{
\citet{Kelly13} present the distribution of Eddington ratios for the Type 1 AGN population, which integrate to 1 over the limited range of the data. The arbitrary rescaling allows us to compare to the probability of a given \emph{galaxy} hosting an AGN, as described by our model C.}}
A steep turnover in the distribution is seen at $\log \lambda\approx-0.5$, indicating the dearth of the highest accretion rate sources amongst the Type 1 AGN population compared to the predictions of our model.}
\amend{Further detailed study of the luminous AGN population, including both Type 1 and Type 2 AGNs, is needed to fully reconcile the differences in mass functions, ERDFs and luminosity functions of the X-ray and optically selected AGN populations.
}

\subsection{The evolution of AGNs}
\label{sec:evol}

Based on our model, we attribute the evolution in the XLF since $z\sim1$ to a reduction in the overall probability of a galaxy hosting an AGN as cosmic time progresses, while the shape of the distribution of accretion rates remains roughly constant. 
This behavior can be interpreted as a rapid drop in the rate of AGN triggering (the duty cycle) or a shift to lower characteristic accretion rates, either of which may be driven by a reduction in the availability of cold gas with cosmic time.
Despite this rapid evolution, AGNs have a wide range of specific accretion rates (with a consistent power-law distribution) and are hosted by galaxies across a wide range of stellar masses at any epoch. 

The details of the physical processes that trigger AGNs and give rise to this distribution of specific accretion rates are still unclear.
The Eddington limit appears to set the maximum rate of SMBH growth at all times (although in rare or short-lived phases this limit may be violated).
Furthermore, the power-law distribution of accretion rates may reflect variability in the level of AGN activity over relatively short timescales (compared to the lifetime of a galaxy), which could also be related to the physical processes in the very central regions of the galaxy that control the rate of SMBH accretion \citep[e.g.][]{Novak11}.
However, variations in larger-scale fueling mechanisms---such as mergers, disk instabilities, cosmological accretion of cold gas, or mass loss from the stellar population of the galaxy \citep[e.g.][]{Kauffmann00,Hopkins06c,Ciotti07}---may also have an impact on the accretion rate distribution.
Indeed, changes in large-scale fueling mechanisms may drive the longer-term evolution that leads to lower characteristic accretion rates or reduced duty cycles at later cosmic times.
The origin of this power-law distribution is is thus a key problem for theoretical studies of AGN physics.

Our model is successful at predicting the predominant evolution of the XLF.
However, we underpredict the number density of moderate luminosity AGNs ($\lx\sim10^{43-44}$ \ergs) at $z\gtrsim0.7$, although the discrepancy is only at the $\sim2\sigma$ level. 
Furthermore, our model lacks the strong, luminosity-dependent evolution favored by most studies of the XLF over wider redshift ranges \citep[out to $z\sim3$, e.g.][]{Ueda03,Hasinger05,Ebrero09,Aird10}.
These luminosity-dependent schemes are associated with ``downsizing" of AGN activity.
Observationally, downsizing is seen as a rapid evolution in the number density of luminous AGNs that peaks at $z\sim2$, while the space density of lower luminosity AGNs evolves more weakly and peaks at lower $z$.
Downsizing may also be reflected in a flattening of the faint-end slope of the XLF, possibly seen in the observed XLF in our highest redshift bin \citep[although ][concluded that a constant shape of the XLF was consistent with the observational data, carefully considering the uncertainties]{Aird10}.
This downsizing behavior is thought to reflect the build up of SMBH masses, whereby the highest mass SMBHs form and grow early in the history of the universe, whereas lower mass SMBHs carry out most of their growth at later times. 
Evidence for this behavior may be seen in the growth rates of AGNs in the local universe \citep[$z<0.3$, e.g.][]{Heckman04}.

Given the success of our model at reproducing the predominant evolution of the XLF at $0.2<z<1.0$, it appears that any downsizing behavior may be a secondary effect at these redshifts. 
Our results indicate that the predominant evolution of the AGN population at $z\lesssim1$ can be associated with a rapid decline in the probability of hosting an AGN for galaxies of any stellar mass as cosmic time progresses. 
In fact, \citet{Aird12} found that the distribution of specific accretion rates was consistent over a wide range of stellar masses---a key assumption of our model---and found no evidence of a downsizing behavior in terms of host stellar mass. 
Nevertheless, the discrepancies between our model and the observed XLF suggest that the specific accretion rate distribution may evolve in a mass-dependent way. Further study is required to confirm such behavior and motivate any refinements to our model.

\subsection{The relationship between AGN activity and star formation}

The evolution of the AGN population has strong parallels with the star formation history of the universe, which may also undergo a downsizing behavior where the most massive galaxies appear to form earlier in the history of the universe \citep[see][for a critical overview of the observational evidence]{Fontanot09}.
However, it is now becoming clear that the rapid decline in the global star formation rate density since $z\sim1-2$ is primarily due to a drop in specific star formation rates for all galaxies, with any stellar-mass-dependent downsizing being a secondary effect \citep[e.g.][]{Noeske07,Zheng07b,Karim11}, similar to our findings for AGNs \citep[see also][]{Georgakakis11b}.
Nevertheless, how and why these processes are connected remains a key open question.

Our simple observationally motivated model assumes that the probability of a galaxy hosting an AGN is determined by a universal specific accretion rate distribution that is independent of host stellar mass \emph{or} star formation properties. 
Thus, the observed AGN population (traced by the XLF) is determined by the galaxy SMF, the scaling between \Mbh\ and \Mstel, and the redshift evolution of the AGN duty cycle. 
However, studies of the correlation between star formation and AGN accretion rates paint a more complex picture, indicating that star formation and AGN accretion rates are connected, at least for the most luminous sources \citep[e.g.][]{Netzer09,Rosario12}, when considering the nuclear regions of galaxies \citep[e.g.][]{Diamond-Stanic12} or when averaged over a large population of galaxies \citep[e.g.][]{Mullaney12b}.
\amend{
Furthermore, \citet{Kauffmann09} studied the distributions of accretion rates for optically selected narrow-line AGNs in the local Universe using the SDSS and found that AGNs have two distinct modes of accretion characterized by very different accretion rate distributions in young, star-forming galaxies and older, passively evolving galaxies. 
Recently, \citet{Rosario13b} also showed that X-ray selected AGNs are more likely to be found in star-forming galaxies on the star-formation main sequence. 
Given these important findings, our model could be improved by considering the SMFs of star-forming and quiescent galaxies separately, with appropriate accretion rate distributions for each population. 
Indeed, \citet{Moustakas13} found that the SMFs of the star-forming and quiescent galaxy populations follow substantially different evolutionary paths.
However, to proceed further with such studies requires accurate measurements of $p(\lambda \giv \mstel,z)$, including any stellar mass and redshift dependence, for each of the galaxy populations (star-forming and quiescent). 
In \citet{Aird12} we split the galaxy population according to optical color and found evidence for a weak (factor $\sim 2$) enhancement in the probability of hosting an AGN for blue galaxies (indicating current star formation) compared to red (quiescent) galaxies, although both populations exhibited a wide, power-law distribution of accretion rates. 
Such relatively minor differences would have relatively little impact on the results of this paper but appear inconsistent with the findings of \citet{Kauffmann09} or \citet{Rosario13b}.
Optical color may be a poor discriminator for the levels of star formation. 
Further study is required to accurately determine how the accretion rate distributions for X-ray AGNs depend on the star formation properties of their host galaxies, refine our observationally motivated scheme to link the galaxy and AGN populations, and fully unveil the interplay between AGN and star formation.}

\subsection{Other models of AGN--galaxy co-evolution}

A number of prior studies have presented models to link the evolution of AGNs and galaxies, either from an observationally motivated standpoint \citep[e.g.][]{Shankar09,Shankar13,Conroy13} or incorporating prescriptions for SMBH growth into either large-scale hydrodynamic cosmological simulations \citep[e.g.][]{Booth09,DeGraf12b} or semi-analytic galaxy formation models \citep[e.g.][]{Bower06,Somerville08,Fanidakis12}. 
Given the current uncertainties regarding SMBH fueling mechanisms and the physics of AGN accretion, the luminosity function is often used to constrain the key parameters in the prescriptions for SMBH growth, such as the AGN duty cycle or the relative rates  of different AGN triggering mechanisms \citep[e.g.][]{Hopkins06b,Shankar13,Conroy13,Draper12}. 
However, up-to-date semi-analytic models, including sophisticated prescriptions for SMBHs, are able to predict the AGN luminosity function \citep[e.g.][]{Fanidakis12,Hirschmann12}, although it can be difficult to assess the level of tuning required for a successful model or robustly test the underlying physical assumptions. 
For example, the \citet{Fanidakis12} and \citet{Hirschmann12} models both reproduce the observed AGN luminosity functions but infer very different levels of AGN triggering due to disk instabilities versus merger-driven inflows.
Furthermore, the \citet{Hirschmann12} model reproduces the observed downsizing trends in the XLF by modifying their fiducial model to include a limit to accretion rates that depends on the cold gas fraction, as well as a prescription for AGN triggering due to disk instabilities and a ``heavy seeding" prescription for the first SMBHs.
\citet{Fanidakis12}, however, attribute the observed downsizing to obscuration effects, incorporated into their model using a prescription based on observations by \citet{Hasinger08}.

Our simple, observationally motivated model takes a very different approach. 
We adopt the observed SMF of galaxies and populate the galaxies with AGNs using models for $p(\lambda \giv \mstel,z)$, based on actual measurements, requiring only a few straightforward assumptions to extend to higher accretion rates. 
Our final model has one free parameter which is tuned to match the observed bright end of the XLF but does not alter our model predictions at lower luminosities. 
Unlike the models described above, our observationally motivated scheme does not provide a complete picture of the relationship between AGNs and their host galaxies or the underlying physical processes that control their co-evolution. 
However, our results indicate that the universal distribution of specific accretion rates, independent of host stellar mass, is the key underlying property of the AGN population that is needed to reproduce the XLF. 
Thus, the distribution of specific accretion rates is a vital observational constraint for theoretical models of AGNs and their evolution within a cosmological context.
Our model also provides a framework to interpret different measurements, understand underlying observational biases that may skew the properties of samples of AGNs, and motivate future observations.
In fact, the lack of downsizing in our model predictions and the discrepancy with the observed XLF for $\lx\approx10^{43-44}$ \ergs\ at $z\sim0.7-1.0$ provides an important motivation for precise measurements of $p(\lambda \giv \mstel,z)$ at these redshifts and higher. 
Ultimately, conclusive tests of cosmological AGN-galaxy models may require comparisons of predictions in the multivariate space of galaxy stellar mass, star formation properties, and the distribution of AGN accretion rates.

\section{Conclusions}
\label{sec:conclusions}

We have developed a simple, observationally motivated model to link the evolution of AGNs and their host galaxies that predicts the X-ray luminosity function of AGNs at $0.2<z<1.0$. 
We use new measurements of the galaxy stellar mass function from PRIMUS \citep{Moustakas13} combined with a simple model to populate these galaxies with AGNs.
Our model is based on our measurements showing that the probability of a galaxy hosting an AGN is defined by a power-law function of specific accretion rate (with slope $\gamma_1\approx0.65$) that depends only on redshift and is independent of host stellar mass \citep{Aird12}.
We extend our model to higher X-ray luminosities and accretion rates by allowing for a break in the specific accretion rate function at a point corresponding to the Eddington limit.
We also allow for scatter in the scaling between SMBH mass and the stellar mass of the host galaxy.
To match the bright end of the XLF, we introduce one free parameter corresponding to the slope of the power-law tail of the distribution of accretion rates above the Eddington limit. 
We constrain this power-law slope to be $\gamma_2=-2.1^{+0.3}_{-0.5}$ by fitting to the observed XLF at $z=0.2-1.0$. 

Our model provides a simple picture of the connection between AGNs and their host galaxies.
Based on the results of \citet{Aird12}, AGN activity may be triggered in galaxies of any stellar mass, with the growth rate of the SMBH being defined by a single distribution of specific accretion rates.
The observed rapid evolution of the AGN population since $z\approx1$ is attributed to an overall reduction in the probability of any galaxy hosting an AGN, indicating a severe reduction in the rate of AGN fueling as cosmic time progresses.
While the physical processes driving this behavior remain unclear, they appear to operate in a uniform manner over a wide range of stellar masses. 
Given our success at reproducing the XLF at $z<1$ with this simple model, 
it appears that the universal distribution of specific accretion rates is the key observational constraint for theoretical models of AGN evolution.

We find that the observed XLF is dominated in number density by moderately massive galaxies ($\mstel\approx10^{11}\msun$), which is due to the combination of the shape of the galaxy SMF and the shape of the specific accretion rate distribution. 
This predominance of moderately massive host galaxies in X-ray selected AGN samples does not represent a preference for AGN activity at any particular stellar mass range.
The observed population of the most luminous AGN may be severely skewed to the most extreme sources with elevated SMBH masses relative to their host galaxies and in short-lived phases of very rapid SMBH accretion.
These sources are likely to provide a biased view of the connection between AGNs and the properties of their host galaxies. 

While our simple model reproduces the predominant evolution of AGN population since $z\sim1$, we do not predict any downsizing in SMBH growth with cosmic time.
We do not fully reproduce the luminosity-dependent evolution in the space density of AGNs or possible changes in the faint-end slope of the XLF at $z\sim0.7-1.0$. 
Thus, the specific accretion rate distribution may evolve in a mass-dependent manner, although any downsizing behavior appears to be a secondary effect at $z\lesssim1$.
Our model also lacks known connections between the star formation in galaxies and the growth of their SMBHs. 
These discrepancies motivate not only the need for high precision measurements of the XLF over a wide range of redshifts but also further detailed observational studies of the distribution of AGN accretion rates within well-defined samples of galaxies divided by stellar mass, redshift, and star formation properties.

\acknowledgements
We thank the referee for helpful comments that have improved this paper.
We acknowledge 
Rebecca Bernstein, Adam Bolton, Scott Burles, Douglas Finkbeiner, David W. Hogg, Timothy McKay, Alexander J. Mendez, Sam Roweis, Wiphu Rujopakarn, Ramin Skibba, and Stephen Smith
for their contributions to the PRIMUS
project. 
We would like to thank the CFHTLS, COSMOS, DLS, and 
SWIRE teams for their public data releases and/or access to early releases.  
This paper includes data gathered with the 6.5 meter Magellan 
Telescopes located at Las Campanas Observatory, Chile.
We thank the support staff at LCO 
for their help during our observations, and we acknowledge
the use of community access through NOAO observing time.
Some of the data used for this project are from the CFHTLS public data release,
which includes observations obtained with MegaPrime/MegaCam, a joint project 
of CFHT and CEA/DAPNIA, at the Canada-France-Hawaii Telescope (CFHT) which is 
operated by the National Research Council (NRC) of Canada, the Institut 
National des Science de l'Univers of the Centre National de la Recherche 
Scientifique (CNRS) of France, and the University of Hawaii. This work is 
based in part on data products produced at TERAPIX and the Canadian Astronomy 
Data Centre as part of the Canada-France-Hawaii Telescope Legacy Survey, a 
collaborative project of NRC and CNRS.
We also thank those who have built and operate the Chandra and XMM-Newton X-ray observatories.
JA acknowledges support from a COFUND Junior Research Fellowship from the Institute of Advanced Study, Durham University. 
This work was also supported by NASA grant NNX12AE23G through the Astrophysics Data Analysis Program.
Funding for PRIMUS has been provided
by NSF grants AST-0607701, 0908246, 0908442,
0908354, and NASA grant 08-ADP08-0019.

\bibliographystyle{apj}

\end{document}